\documentclass[doublecol]{epl2} 

\usepackage[english]{babel}
\usepackage{graphicx}
\usepackage{dcolumn}
\usepackage{bm}
\usepackage{color}
\usepackage{sidecap}
\usepackage{subfigure}

\newcommand{\zbar}{\bar{z}}
\newcommand{\dzbar}{\partial_{\bar{z}}}
\newcommand{\Vbar}{\bar{V}}
\newcommand{\cbar}{\bar{\chi}}
\newcommand{\ebar}{\bar{\epsilon}}

\title{Realization of the driven nonlinear Schr\"odinger equation with stationary light}

\author{P. Das\inst{1} \and C. Noh\inst{1} \and D. G. Angelakis\inst{1,2}}
\shortauthor{P. Das \etal}

\institute{                    
  \inst{1} Centre for Quantum Technologies, National University of Singapore, 3 Science Drive 2, Singapore 117543\\
  \inst{2} Science Department, Technical University of Crete, Chania, Crete, Greece, 73100
}

\abstract{
We introduce a versatile platform for studying nonlinear out-of-equilibrium physics. The platform is based on a slow light setup where an optical waveguide is interfaced with cold atoms to realize the driven nonlinear Schr\"odinger equation with a potential. We compare the proposed setup with similar setups using Bose-Einstein condensates and investigate the system's response under coherent driving for a lattice potential. The slow light setup provides novel angles in the study of nonlinear dynamics due to its advantages in introducing and modulating the driving, the extra tunability over the sign and strength of the available nonlinearities, and the possibility to electromagnetically carve out the underlying potential on demand.}

\begin{document}

\maketitle

\section{Introduction}
The nonlinear Schr\"odinger equation (NLSE) arises naturally in many areas of physics. It describes the propagation of an electromagnetic field in a nonlinear medium \cite{NOP0,NOP1} or the interacting ground state of a Bose-Einstein condensate where it is known as the Gross-Pitaevskii equation \cite{Pitaevskii}. The out-of-equilibrium physics in such systems are of fundamental importance, where the nonlinearity of the system changes transport properties of a driven system, for example. Previous studies under this category include the nonlinear transport properties of Bose-Einstein condensates (BECs) in various potentials \cite{Carusotto,BNT0,BNT1,BNT2,Morsch,Raizen,Zenesini,Fabre,Aspect} and soliton propagation in nonlinear dielectrics \cite{NOP2}.  Quantum transport of few-quanta pulses in a strongly nonlinear electromagnetically induced transparency (EIT) setup \cite{Schmidt} has also been studied recently, aimed at photon switching applications \cite{HCT1,HCT5}.

In this letter, we introduce a versatile platform for studying nonlinear out-of-equilibrium physics, based on the stationary light system \cite{STL0,STL1}. The ability to controllably introduce a tunable potential for a trapped correlated polariton gas \cite{HCT4}, makes it an attractive platform to realize and explore quantum nonlinear dynamics complementing cold atoms and nonlinear optics setups. 
As an example of out-of-equilibrium physics, the transport problem is investigated in the semiclassical limit, where we assume that a coherent input field is continuously driving the waveguide. In this case, the resulting dynamics for the trapped fields are of the driven NLSE nature and the transmission spectrum in the linear and nonlinear regimes can be studied.   In the linear case the setup realizes a textbook transmission problem where the transmission resonances and the bandgap due to periodicity are observed. In the nonlinear case, the interplay between the nonlinearity and the lattice potential is manifest. We further perform a linear stability analysis to investigate the feasibility of observing the transmission spectra. Our analysis reveals that only the lowest branch of the transmission spectra in the multi-valued region will be populated. We briefly mention a possible method to stabilize other branches.

\begin{figure}[t]
\includegraphics[scale = 0.42]{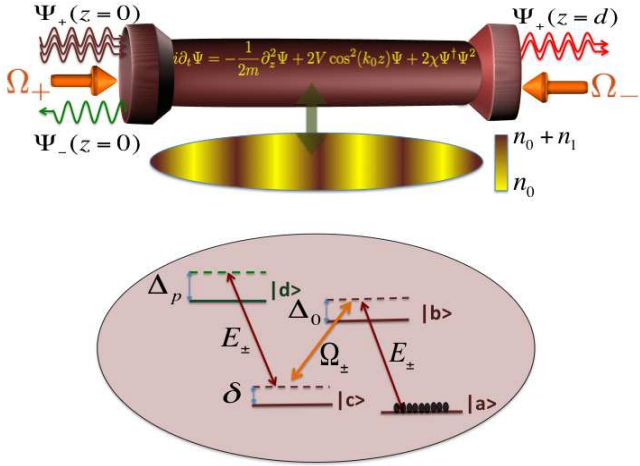}
\caption{Schematic diagram of the system consisting of an optical waveguide interfaced with an ensemble of cold atoms. Two counter propagating control fields, $\Omega_\pm$, drive the 4 level atoms and create a Bragg grating that traps the incident light $\Psi_{+}$. The modulation in the atomic density induces an effective periodic potential acting on the trapped light. The trapped stationary dark-state polaritons  formed in the waveguide can be made to obey the nonlinear Schr\"odinger equation with optically tunable interaction parameters.
}
\label{schematic}
\end{figure}

\section{System}
Figure \ref{schematic} depicts a schematic diagram of our proposed setup, an optical waveguide  interfaced with an ensemble of cold atoms, and the EIT-based atomic level configuration. We note that possible experimental platforms can be based on a hollow core fiber filled with ultracold atoms  \cite{HCE1,HCE3,HCE4,HCE5,HCE6} or a tapered fiber evanescently coupled to cold atoms trapped near the surface of the fiber \cite{TFE1,TFE2}. The resulting dynamics is formally similar to that of a BEC in an optical lattice and also systems studied in nonlinear optics. Our system, however, exhibits advantages such as more efficient preparation of the source (drive), the ability to control the (effective) mass and the (EIT induced) interaction parameters over a wide range of values. Finally, we note that the proposed system also allows the study of out-of-equilibrium quantum many body physics, which we plan to investigate in the future.

As described in \cite{HCT1,HCT5,Fleischhauer,Unanyan}, the dark-state polariton operators describing the trapped fields are $\hat{\Psi}_{\pm} = g \sqrt{2 \pi n_{0}} \hat{E}_{\pm} / \Omega_{\pm}$, where $\hat{E}_{\pm}$ are the slowly-varying parts of the trapped quantum fields and $\Omega_{\pm} = \Omega$ are classical control fields with $\pm$ denoting the directions the fields are travelling in; $n_{0}$ is the mean density of atoms coupled to the waveguide and $g$ is the atom-field coupling strength. Introducing the symmetric and anti-symmetric combinations of the polariton operators $\Psi = (\Psi_{+} + \Psi_{-}) / \sqrt{2}$ and $A = (\Psi_{+} - \Psi_{-})/\sqrt{2}$, $A$ can be shown to adiabatically follow $\Psi$, i.e., $A = - i k_{0} L_{coh} (m_{R}/m) \partial_{z} \Psi$, in the limit of large optical depth. $k_{0}$ is the wave number of the lattice potential and $L_{coh} = \frac{|\Delta_{0}|^{2} + (\Gamma/2)^{2}}{\Gamma_{1D} n_{0} |\Delta_{0}|}$ is the characteristic length scale, termed the coherence length \cite{HCT5}; $m = - \frac{\Gamma_{1D} n_0}{4 \nu_g \left(\Delta_0 + i \Gamma/2 \right)}$  is the complex effective mass and $m_R$ is the real part of $m$; $\nu_g \approx \nu\Omega^2/(\pi g^2 n_0)$ is the group velocity with $\nu$ the velocity in an undoped fibre; $\Delta_{0}$ is the two photon detuning shown in Fig.~\ref{schematic}; $\Gamma_{1D}$ denotes the sponteneous emission rate into the guided modes; $\Gamma$ is the total spontaneous emission from the excited states $|b\rangle$ and $|d\rangle$.

A polaritonic lattice potential can be introduced by creating a periodic modulation in the atomic density: $n = n_0 + n_1 \cos^2(k_0z)$ with $n_1 \ll n_0$ \cite{HCT4}. In this case the dynamics of the symmetric polariton operator is governed by the NLSE: 
\begin{equation}
\label{LL}
i\partial_t \Psi  = - \frac{1}{2m} \partial_z^2\Psi + 2 V \cos^2(k_0 z) \Psi + 2 \chi \Psi^\dagger\Psi^2,
\end{equation}
where $V$ is an effective lattice depth and $\chi$ an effective polariton-polariton interaction strength. Substituting $\Psi_{\pm}(z,t) = \psi_{\pm}(z,\epsilon) e^{- i \epsilon t}$ into the above equation yields the following dimensionless coupled mode equations:
\begin{eqnarray}
\label{CDE1}
\dzbar \psi_+ &-& \frac{i}{2}\frac{m}{l_{coh} m_\textrm{R}} \left( \psi_+ - \psi_-\right) - \frac{i l_{coh}}{2} \left[ \ebar -  \Vbar - \Vbar \cos(2\zbar) \right. 
\nonumber \\  & -& \left.  \frac{\cbar}{2} (\psi_+ + \psi_-)^\dagger (\psi_+ + \psi_-) \right] ( \psi_+ + \psi_- ) = 0, \\
\dzbar \psi_- &-& \frac{i}{2}\frac{m}{l_{coh} m_\textrm{R}} \left( \psi_+ - \psi_-\right) + \frac{i l_{coh}}{2} \left[ \ebar -  \Vbar - \Vbar \cos(2\zbar) \right. 
\nonumber \\  & -& \left.  \frac{\cbar}{2}\left(\psi_+ + \psi_-\right)^\dagger\left(\psi_+ + \psi_-\right)\right]\left( \psi_+ + \psi_- \right) = 0,
\label{CDE2}
\end{eqnarray}
where $z$ and $\psi_\pm$ have been scaled by $1/k_{0}$, $\sqrt{k_{0}}$ and other effective interaction parameters by the recoil energy $E_{R} = k^{2}_{0}/(2 m_{R})$, except $\chi$ which has been further scaled by $1/k_0$; $l_{coh} = k_{0} L_{coh}$ quantifies the ratio between the coherence length and the lattice constant. We note in passing that the above coupled mode equations are similar, but not equivalent, to the nonlinear coupled mode equations arising in the study of nonlinear wave propagation in one-dimensional periodic structures \cite{NOP3}.

\begin{figure}[t]
\vspace{0.5cm}
\includegraphics[scale=0.23]{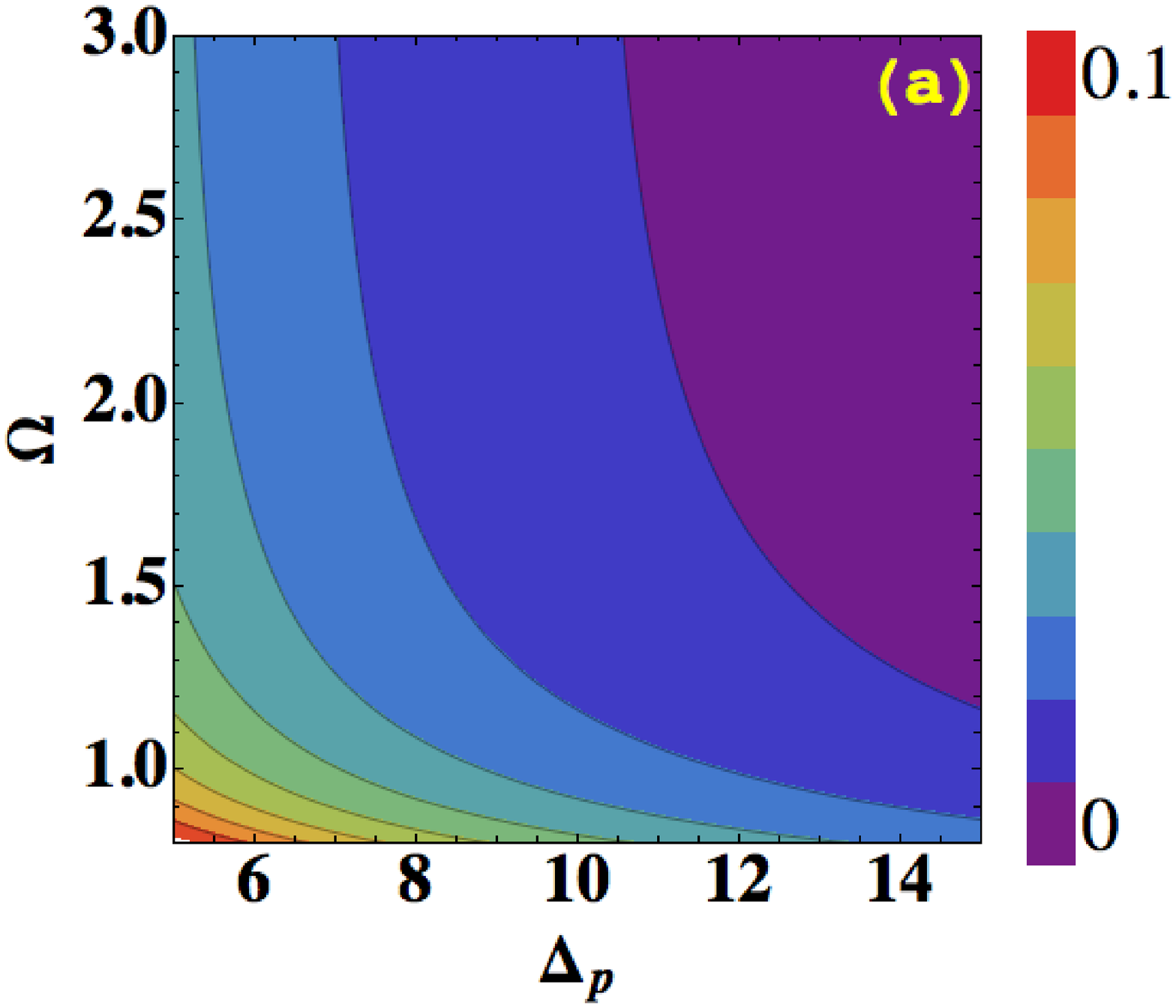}
\includegraphics[width= 3.8cm,height=3.9cm]{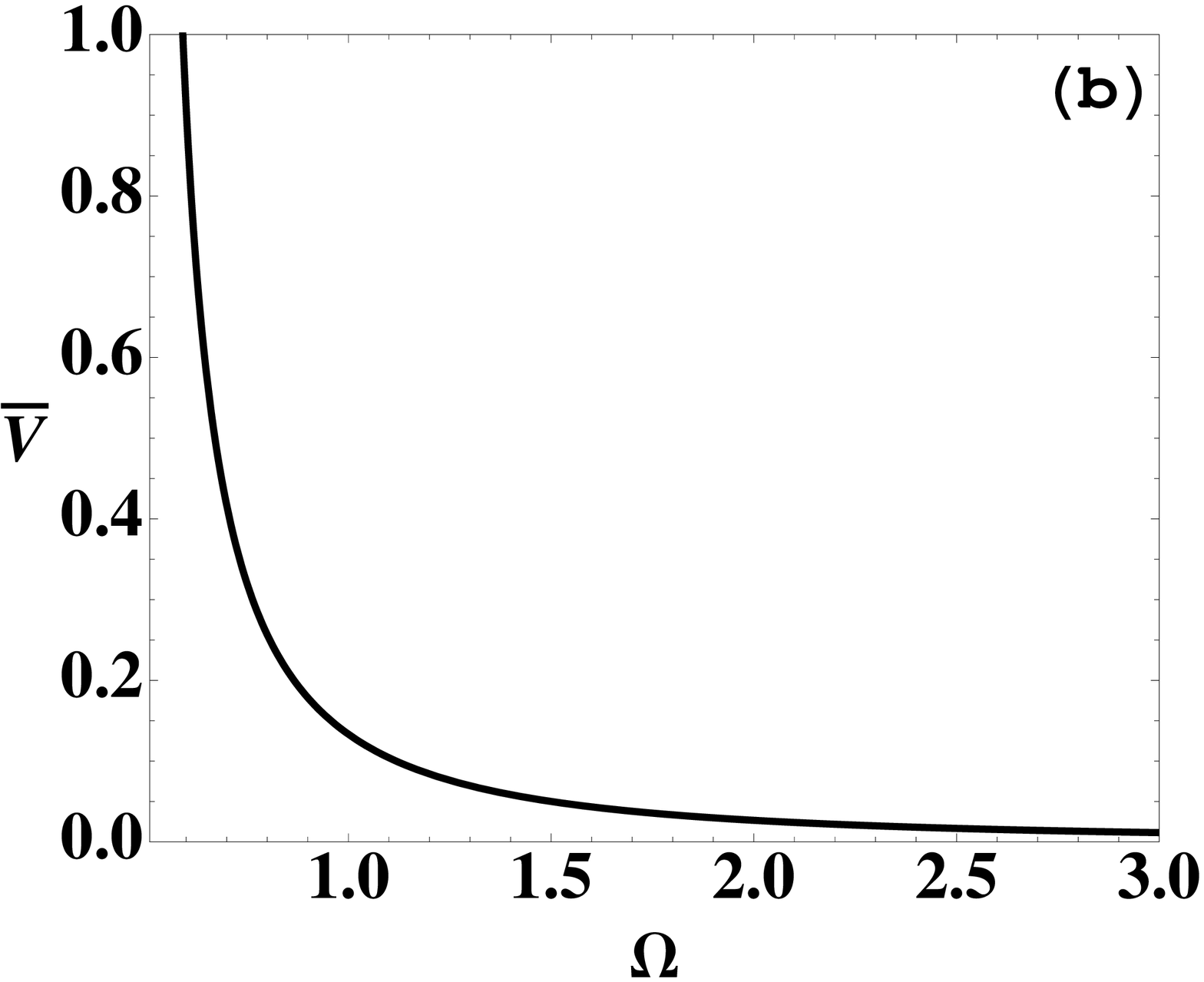}
\caption{(a) The real part of the interaction parameter $\cbar$ as a function of one photon detuning $\Delta_{p}/\Gamma$ and Rabi frequency $\Omega/\Gamma$. (b) The effective polaritonic lattice depth as a function of Rabi frequency $\Omega/\Gamma$. } 
\label{prange}
\end{figure}

In terms of bare optical parameters, the dimensionless lattice depth and nonlinearity are given by: $\Vbar = \frac{\Lambda\Gamma_{1D}^2 \delta n_0 n_1 |\Delta_0|}{8\Omega^2 k_0^2 (\Delta_0^2 + \Gamma^2/4)}$ and $\cbar = \frac{\Lambda^2\Gamma_{1D}\Delta_p}{4l_{coh}(\Delta_p^2+\Gamma^2/4)}\left(1-i\frac{\Gamma}{2\Delta_p}\right)$, where $\Lambda = \Omega^2/(\Omega^2-\delta\Delta_0/2)$. $\Delta_{p}$ and $\delta$ are the single and two photon detunings shown in Fig.~\ref{schematic}. For the optical parameters, we assume $\delta = -0.01$, $\Delta_0 = -50$, $\Gamma_{1D} = 0.2$, $n_0 = 10^{7} m^{-1}$, $n_1 = 10^{6} m^{-1}$, $k_0= 10^{4} m^{-1}$, where all the frequencies are in units of the spontaneous emission rate, $\Gamma$. With these values, which are within reach of near future experiments in tapered and hollow core fibers \cite{HCE1,HCE3,HCE4,HCE5,HCE6,TFE1,TFE2}, a broad range of effective interaction parameters can be obtained as depicted in Fig.~\ref{prange}. 

Equation (\ref{LL}) is equivalent to the Gross-Pitaevskii equation with an optical lattice potential, showing formal similarity between the proposed photonic system and a BEC-based system. While transport problems involving cold atoms and BECs have been studied extensively \cite{Raizen,Carusotto,BNT0,BNT1,BNT2,Morsch,Zenesini,Fabre,Aspect}, our proposal utilizes a fundamentally different system where the tunably-interacting fields are hybrid light-matter excitations and therefore offer a complementary system with distinct advantages.
One important advantage is that the non-equilibrium driving conditions arise much more naturally in the present system. Preparing a coherent source of an atomic BEC is much more difficult compared to preparing a coherent source of photons, i.e., a laser, although there have been a great progress in producing a guided matter waves \cite{Fabre,Couvert,Dall,Cheiney}. The proposed system also allows efficient measurements of the output intensities and correlation functions using standard optical methods, which are difficult to perform in experiments involving BECs.

In this letter, we investigate the transport dynamics in the situation depicted in Fig.~\ref{schematic}, where an input field impinges on the waveguide from the left. We specify the natural boundary conditions $\psi_{+}(z=0) = \alpha$ and $\psi_{-}(z=d) = 0$, where $\alpha$ depends on the field strength of the driving laser impinging from the left and $d = k_{0} L$ is the dimensionless length of the waveguide. The transmission spectrum, i.e., the transmittivity as a function of the probe field detuning $\ebar$, is investigated in both the linear and nonlinear regimes in the semiclassical limit with emphasis on effects of the lattice potential. Finally, a linear stability analysis is performed to show the prospect of observing the characteristic features found in the transmission spectrum.

\section{Linear regime}
In the linear regime, our system provides a realization of a classic transmission problem of a particle moving through a potential. Thus, using the periodic optical medium, it is possible to build a tunable optical experiment to simulate the Schr\"odinger equation in a controlled environment using stationary photons. Moreover, the shape of the photonic potential can be engineered, by carving out the atomic density distribution as shown in \cite{HCT4}, to test other classic text-book problems. Note that the regime can be achieved without the 4th level in Fig.~\ref{schematic}, which alleviates experimental difficulties considerably. 

\begin{figure}[ht]
\includegraphics[width= 4cm]{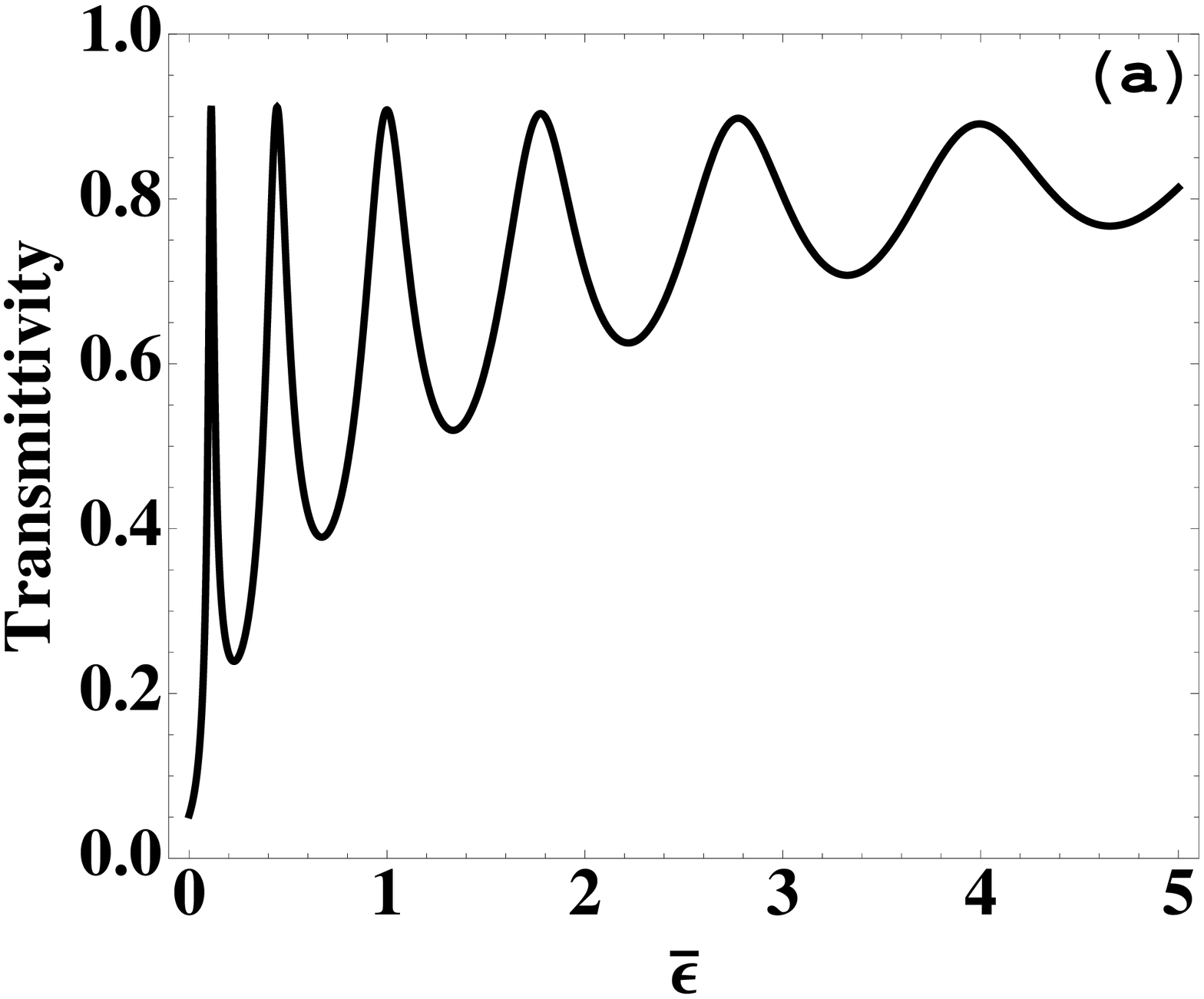}
\includegraphics[width= 4cm]{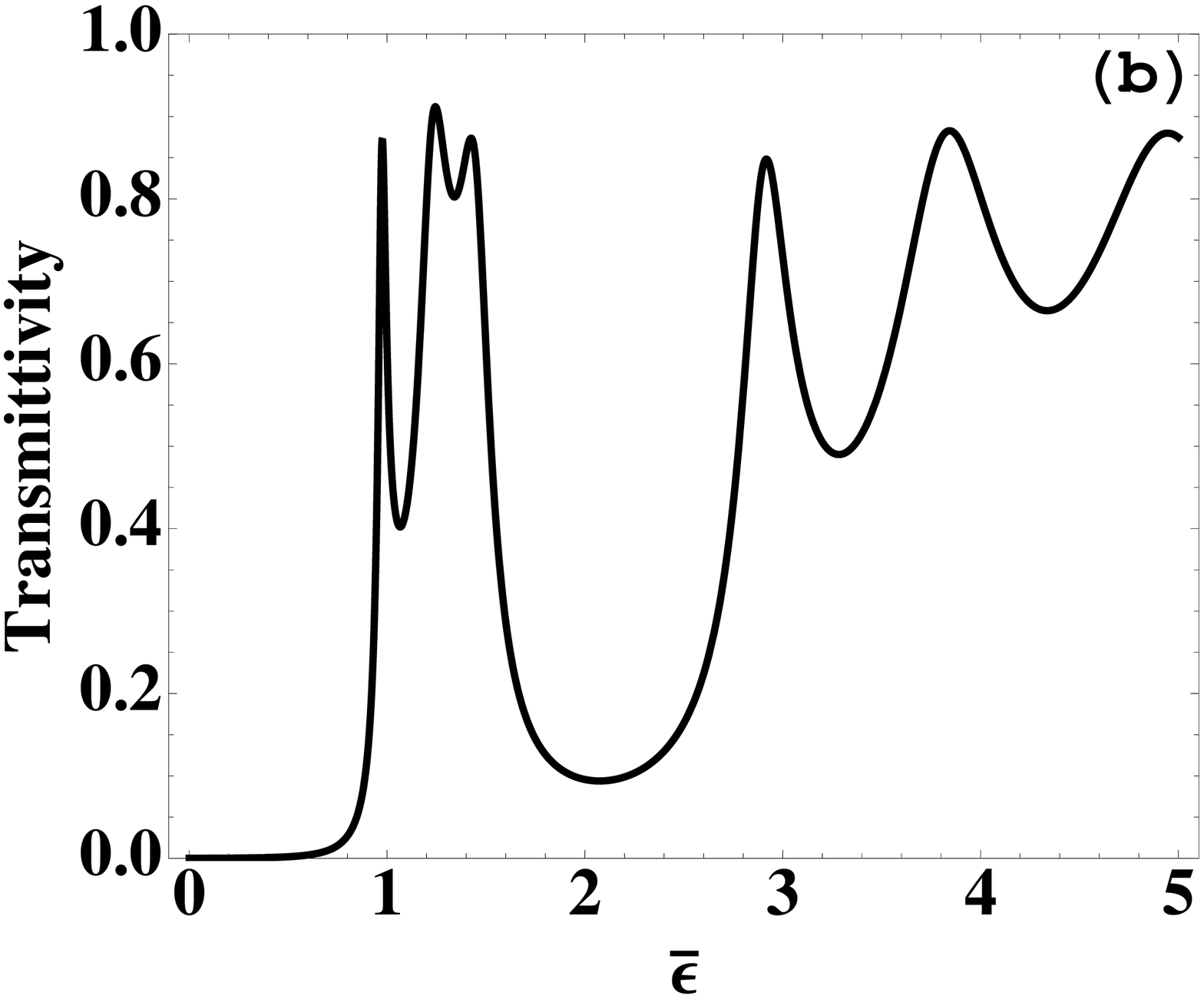}
\caption{Transmission spectra for $l_{coh} = 0.25$, $d=3 \pi$, $\beta = 12\pi/50 \approx 0.75$, $\Vbar = 0$ (a),  and $\Vbar = 1.0$ (b). The bound state and the first band gap is clearly shown in (b).}
\label{fig:linear1}
\end{figure}

Figure \ref{fig:linear1} shows the transmittivity as a function of the input field detuning for two different values of $\Vbar$ for $l_{coh} = 0.25$ with the total loss determined by $\beta = (d/l_{coh})(\Gamma/|\Delta_0|) = 12\pi/50 \approx 0.75$.  

When $\Vbar=0$, i.e., when $\Omega/\Gamma$ is large enough, there are transmission peaks at the resonances $\ebar_0 = (n\pi)^2$ for $n=1,2,3,...$ and at an additional point $\ebar_0 = 1$. These transmission resonances are an artefact of the Bragg trapping produced by the interaction of the counter-propagating control fields with the atoms \cite{HCT5}. As $\Omega/\Gamma$ is tuned down to increase the effective potential depth to 1, the continuum shifts towards right by $2\Vbar$ (maximum potential height) as expected, and additional resonances develop in $\ebar < 2\Vbar$ signifying bound states. The energy of the first bound state can be readily calculated within the perturbation theory as $\Vbar-\Vbar^2/8$, which matches the starting point of the first resonance in the figure. Furthermore, the locations of other resonances can be approximated from a generalization of the free field resonance points $d\sqrt{\ebar} = n\pi$ to $d\sqrt{\ebar-\Vbar+\Vbar^2/8} = n\pi$, such that only the quasi-kinetic energy part is used to determine the resonance condition. This approximation works well if the input field detuning is much larger than the effective lattice depth, i.e. $\ebar \gg \Vbar$, because in this case the effects of the potential is perturbative and the solution is expected to be close to the free field case. 

\section{Non-linear regime}
The nonlinear transport problem in the presence of a potential has been studied in detail in various physical systems. As we have explained earlier, however, the present system offers unique advantages over the previous works: tunability of the parameters, large nonlinearity and readily available coherent sources. We focus on the transmission spectrum as an example of the transport problem and study the interplay between the nonlinearity and periodicity. 

\begin{figure}[ht]
\includegraphics[width= 4cm]{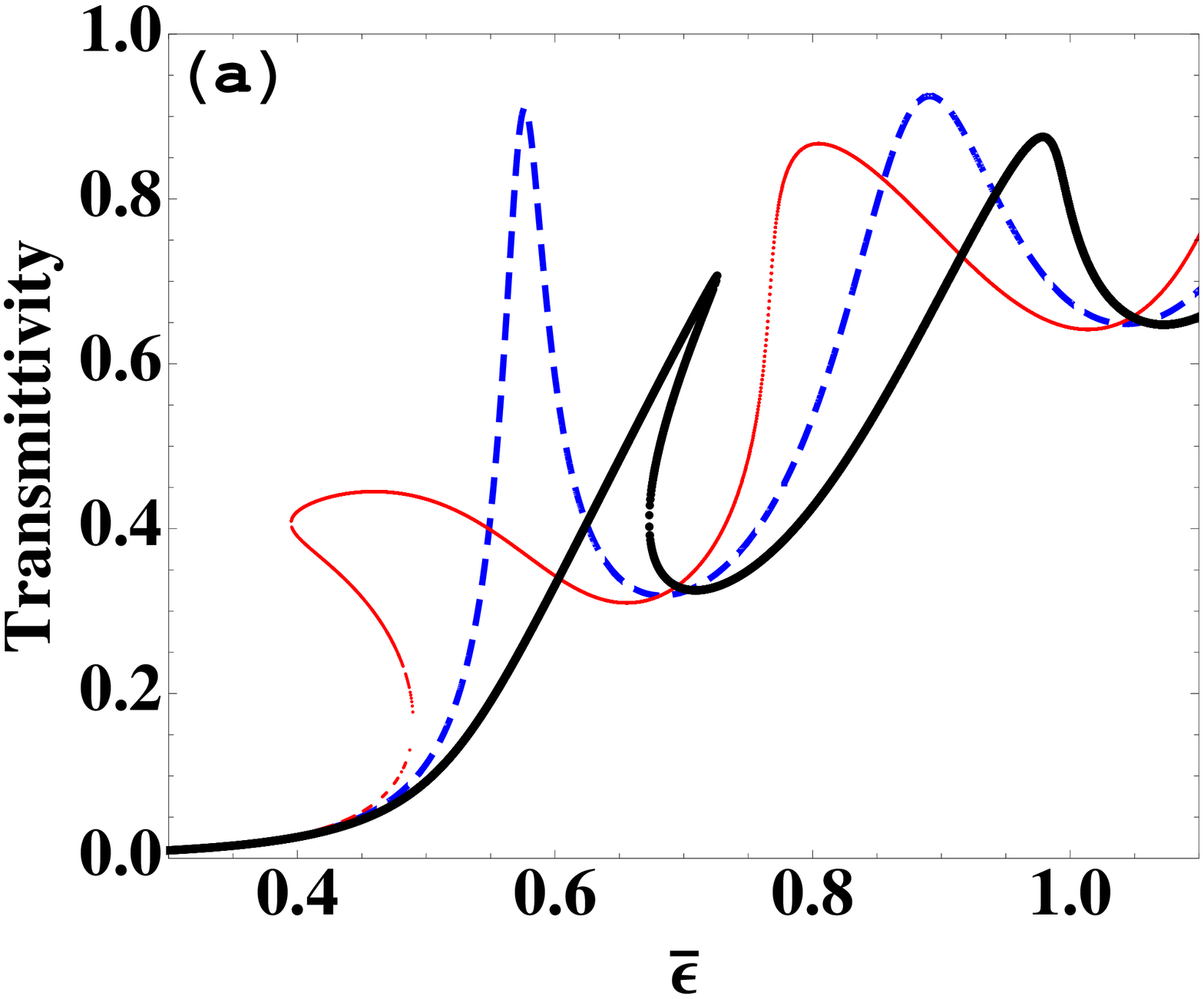}
\includegraphics[width= 4.3cm]{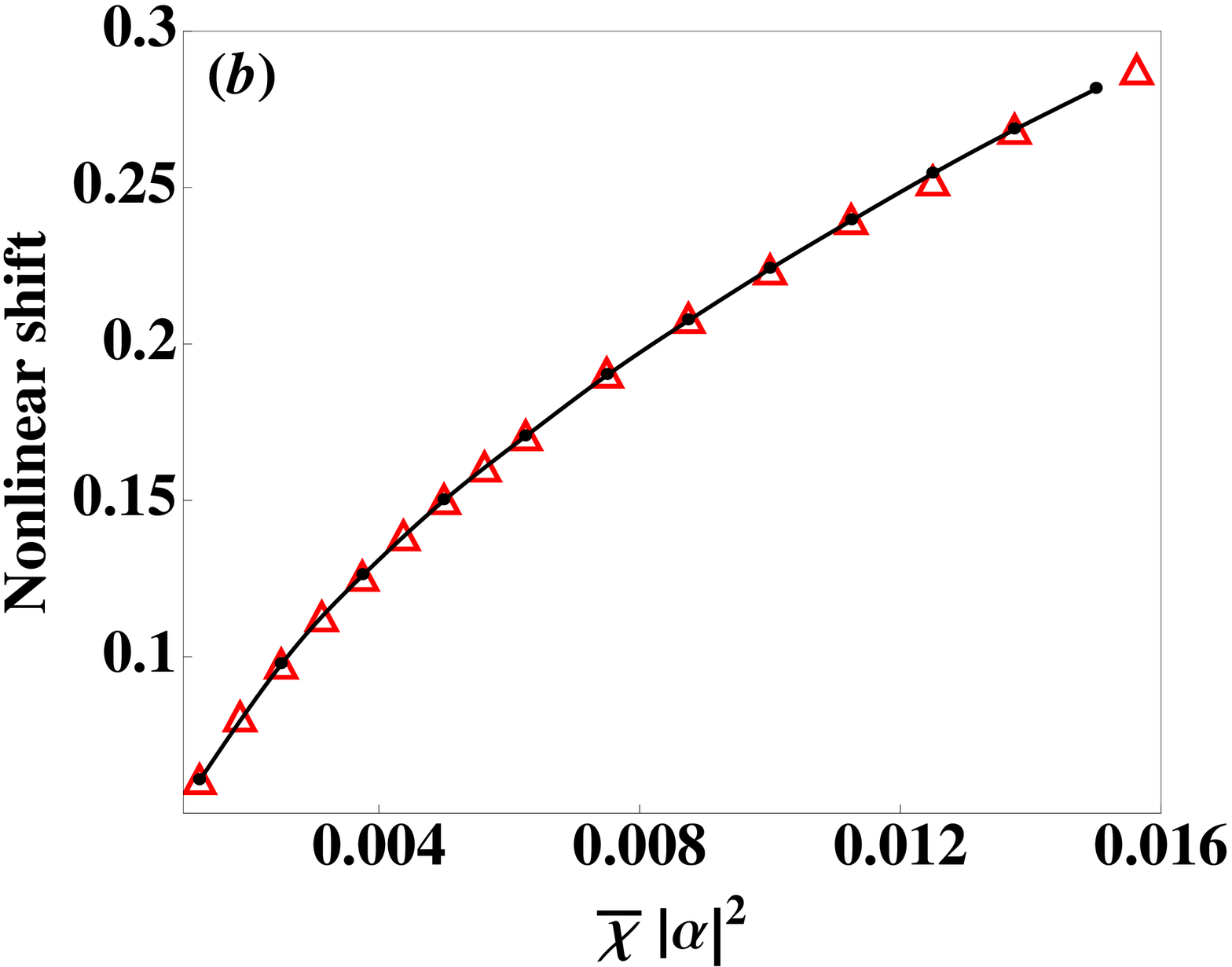}
\caption{(a) Transmission spectra for $\Vbar=0.5$, $l_{coh} = 0.25$, $\alpha = 0.5$, $d = 3 \pi$, $\beta \approx 0.75$, and three different values of $|\cbar|$. For repulsive nonlinearity ($\cbar = 0.02$, $\Delta_p > 0$), one can see the shift of the transmission peaks to the right (thick, black curve). For attractive nonlinearity ($\cbar = - 0.02$, $\Delta_p < 0$), the shift is in the opposite direction (thin, red curve). (b) Nonlinear shift as a function of the effective nonlinearity, $\cbar|\alpha|^2$, for different values of $|\alpha|$. The solid line (black) and the triangular points (red) correspond to $\alpha = 0.5$ and $0.25$, respectively.}
\label{fig:nonlinear}
\end{figure}

As we tune the single photon detuning $\Delta_p$ from a positive to a negative value, the self nonlinearity changes and the magnitude and the direction of the nonlieanr shift changes. This is depicted in Fig.~\ref{fig:nonlinear}(a) where the transmission spectra for $\cbar = -0.02,0,0.02$ are shown by the thin (red), dashed (blue) and thick (black) curves respectively. Higher magnitudes of nonlinearity, corresponding to smaller $\Delta_p$, result in lower transmission peaks due to higher nonlinear losses. The sign and magnitude of $\cbar$ can be tuned by varying, for example, the single-photon detuning $\Delta_p$ as shown in Fig.~\ref{prange}(a).

Fixing $\cbar$ and increasing the input field strength $\alpha$ has similar effects on the transmission spectrum to fixing $\alpha$ and increasing $\cbar$. This is expected because the effective photonic nonlinearity inducing the nonlinear shift is determined by $\cbar|\bar{\psi}|^2$, where the overbar denotes averaging. $|\bar{\psi}|^2$ of course depends on the value of the driving coherent field amplitude $\alpha$, but because of the nonlinearity, the exact nature of this dependence is not obvious. To understand the physics better, we have investigated the amount of shift of the first transmission peak, with respect to the linear case, as a function of the effective nonlinearity, $\cbar |\alpha|^{2}$. The result is depicted in Fig.~\ref{fig:nonlinear}(b), where the solid line (black) represents the shift for $\alpha = 0.5$ and the triangular points (red) correspond to $\alpha = 0.25$. The nonlinear shift depends only on the effective nonlinearity to a very good approximation and furthermore, the dependence seems to be almost linear for larger values of the effective nonlinearity. Therefore, one should be able to observe similar effects for even lower values of nonlinearity.

\begin{figure}[ht]
\includegraphics[width= 4cm]{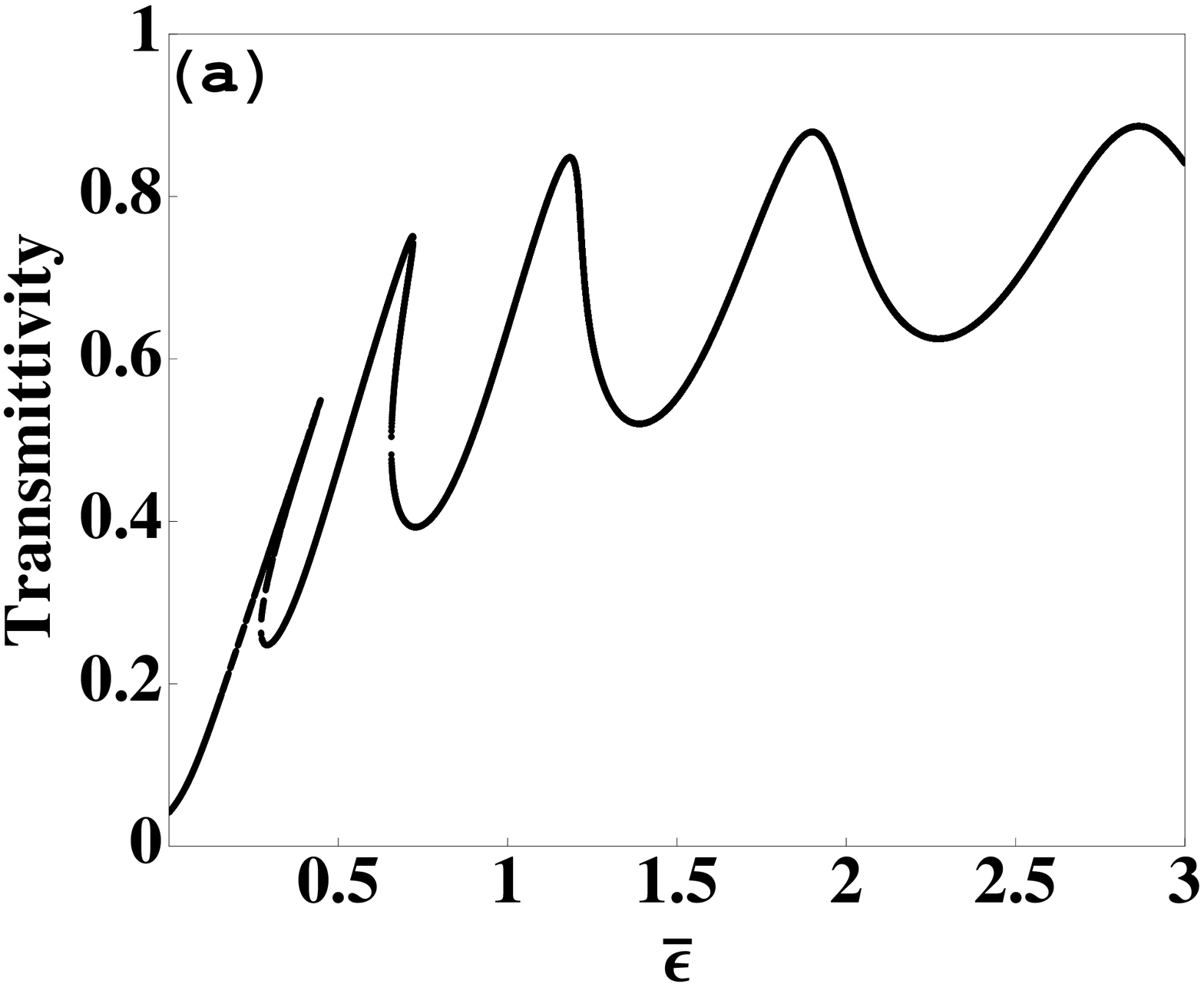}
\includegraphics[width= 4cm]{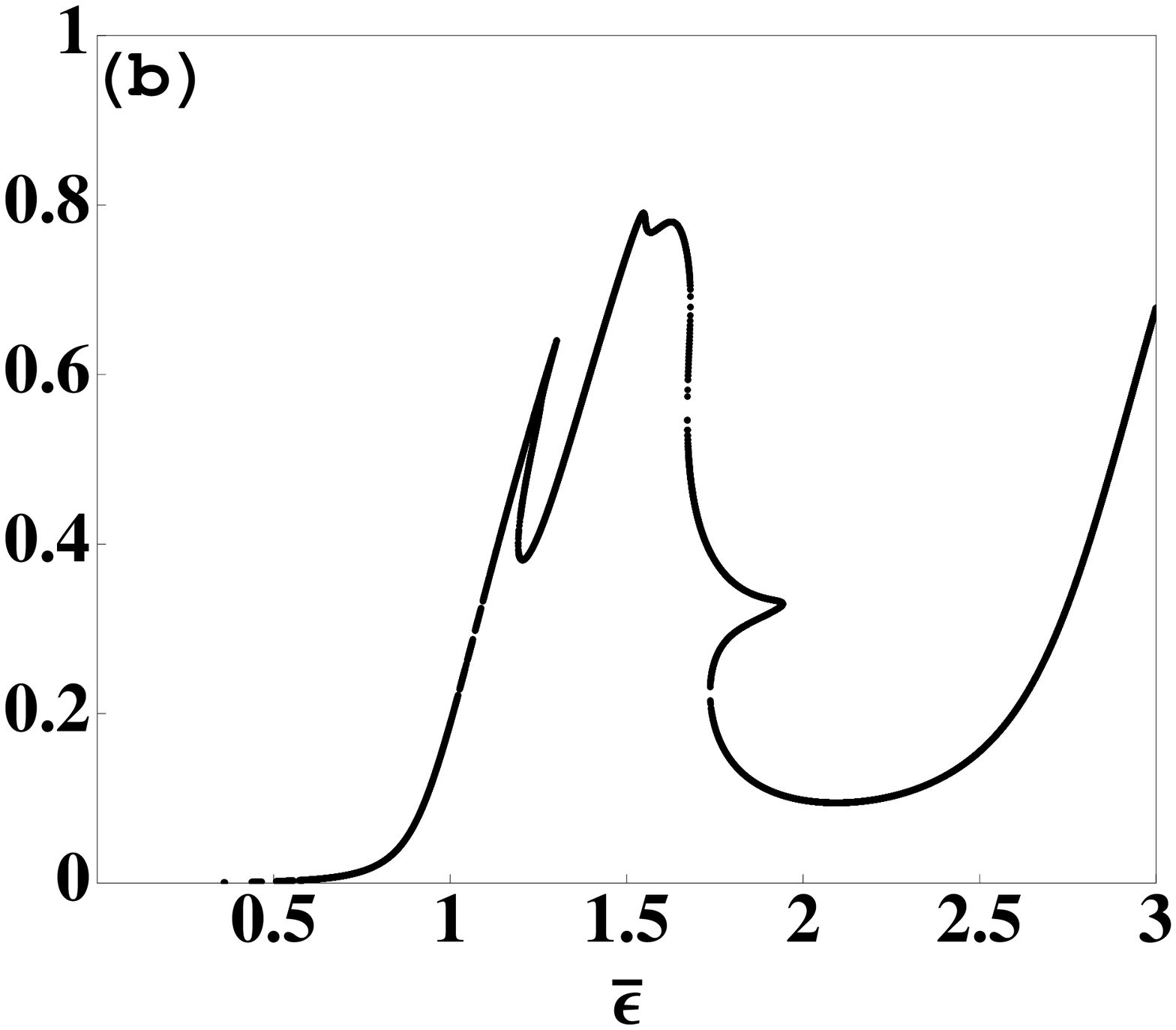}
\caption{Transmission spectra for $l_{coh} = 0.25$, $\alpha = 1.0$, $\cbar=0.02$, $d = 3 \pi$, $\beta \approx 0.75$, and $\Vbar = 0$ (a) and $1$ (b).}
\label{varying_V}
\end{figure}

\begin{figure}[ht]
\centering
\includegraphics[width= 4.5cm]{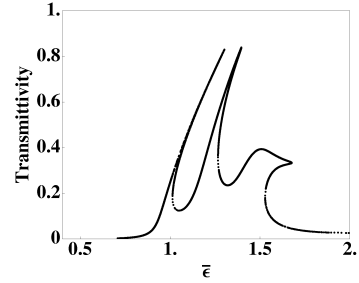}
\caption{Appearance of a `duck-head'-like feature for small coherence length. The parameter values are: $l_{coh} = 0.1$, $\alpha = 0.1$, $\cbar=0.1$, $d = 3 \pi$, $\beta  \approx 0.94$, and $\Vbar = 1$.}
\label{duck-head}
\end{figure}

Next, we investigate the effects of the polaritonic lattice on the transmission properties of the photons in the repulsive case, $\Delta_p >0$. As we decrease $\Omega$, the entire spectrum shifts to the right as shown in Fig.~\ref{varying_V}. At the same time, a gap develops where the transmission is strongly suppressed. The result of the shift and the formation of the bandgap is that the resonance peaks get pushed towards each other as the polaritonic lattice deepens. Near $\Vbar = 1.0$, the second and the third peak merges and a peculiar feature is observed between $\ebar = 1.5$ and $\ebar = 2.0$: a horizontal peak near the bandgap region appears, as seen in Fig.~\ref{varying_V}(b). As far as we are aware, this feature has not been observed before and is a unique feature arising due to the interplay between the the nonlinear-shift and the band gap. It is easier to see the origin of this feature with more extreme parameters as shown in Fig.~\ref{duck-head}. A `duck head'-like feature is observed, which arises due to stronger suppression of the transmission at the tip of the transmission resonance that falls on the edge of the bandgap. As $l_{coh}$ is increased, the resonance peaks broaden and part of the `duck head' structure gets absorbed, resulting in the horizontal peak shown in Fig.~\ref{varying_V}(b).

\begin{figure*}[t]
\centering
\subfigure[$ $]{
    \includegraphics[scale= 0.22]{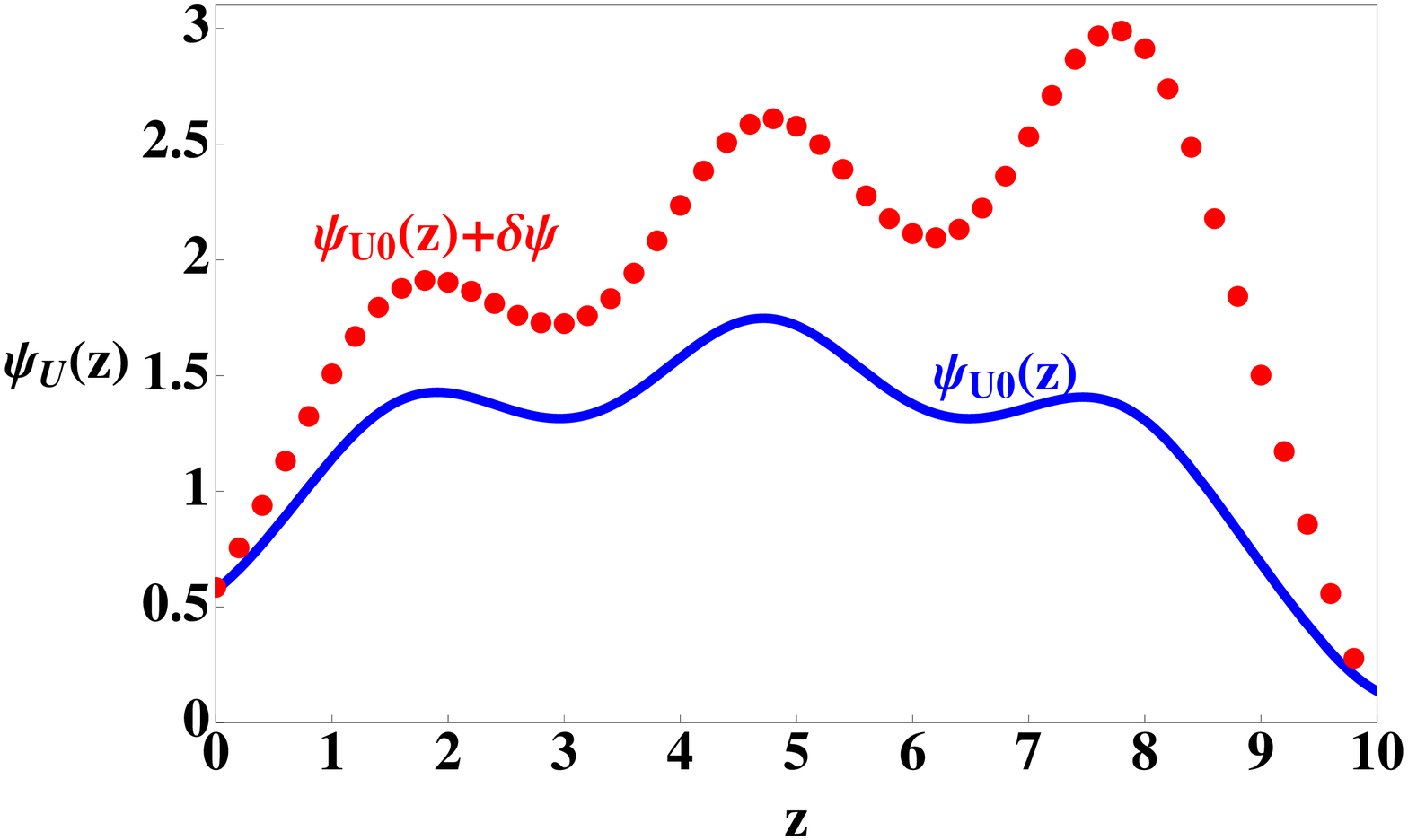} }
\subfigure[$ $]{
    \includegraphics[scale= 0.22]{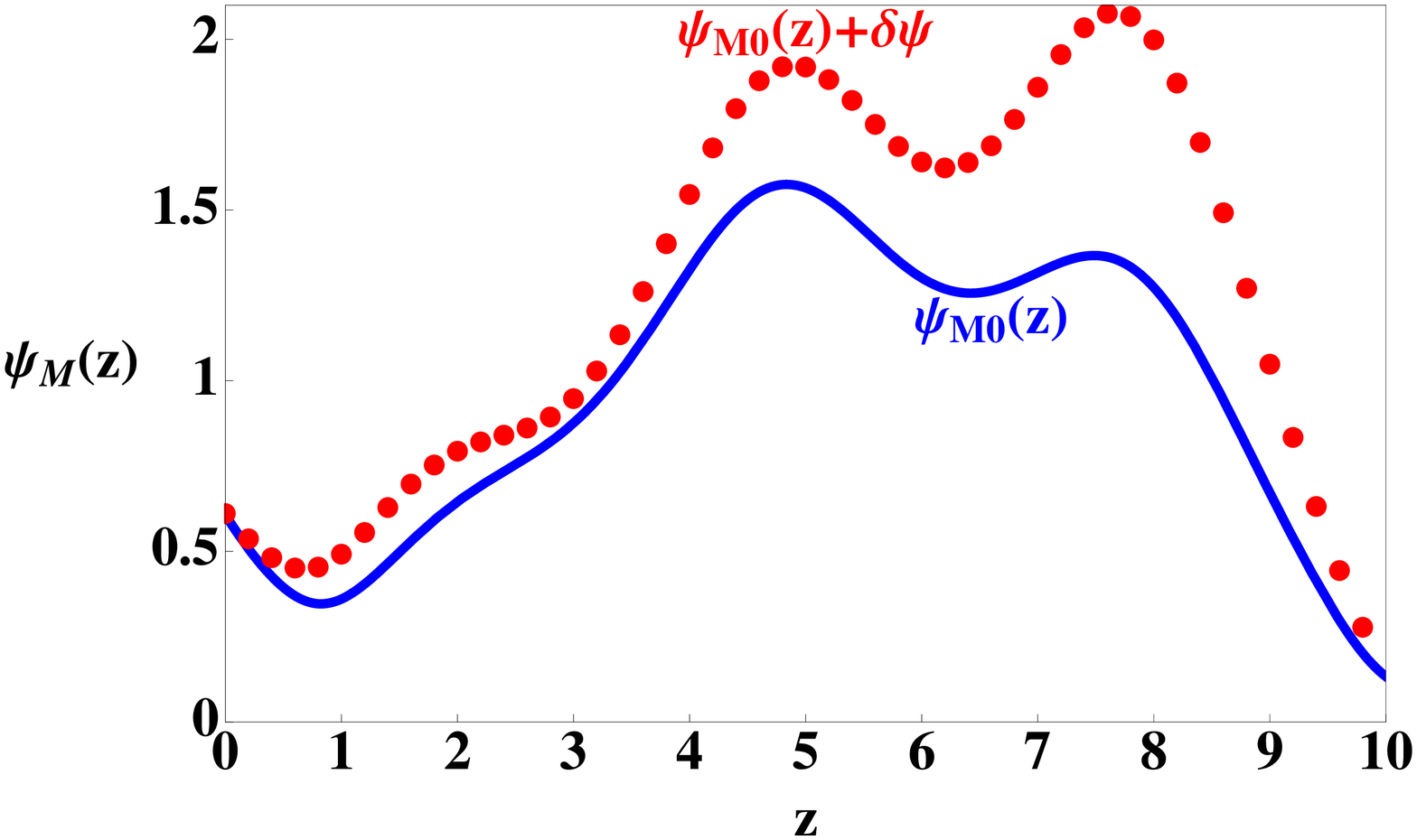}}
\subfigure[$ $]{
    \includegraphics[scale= 0.22]{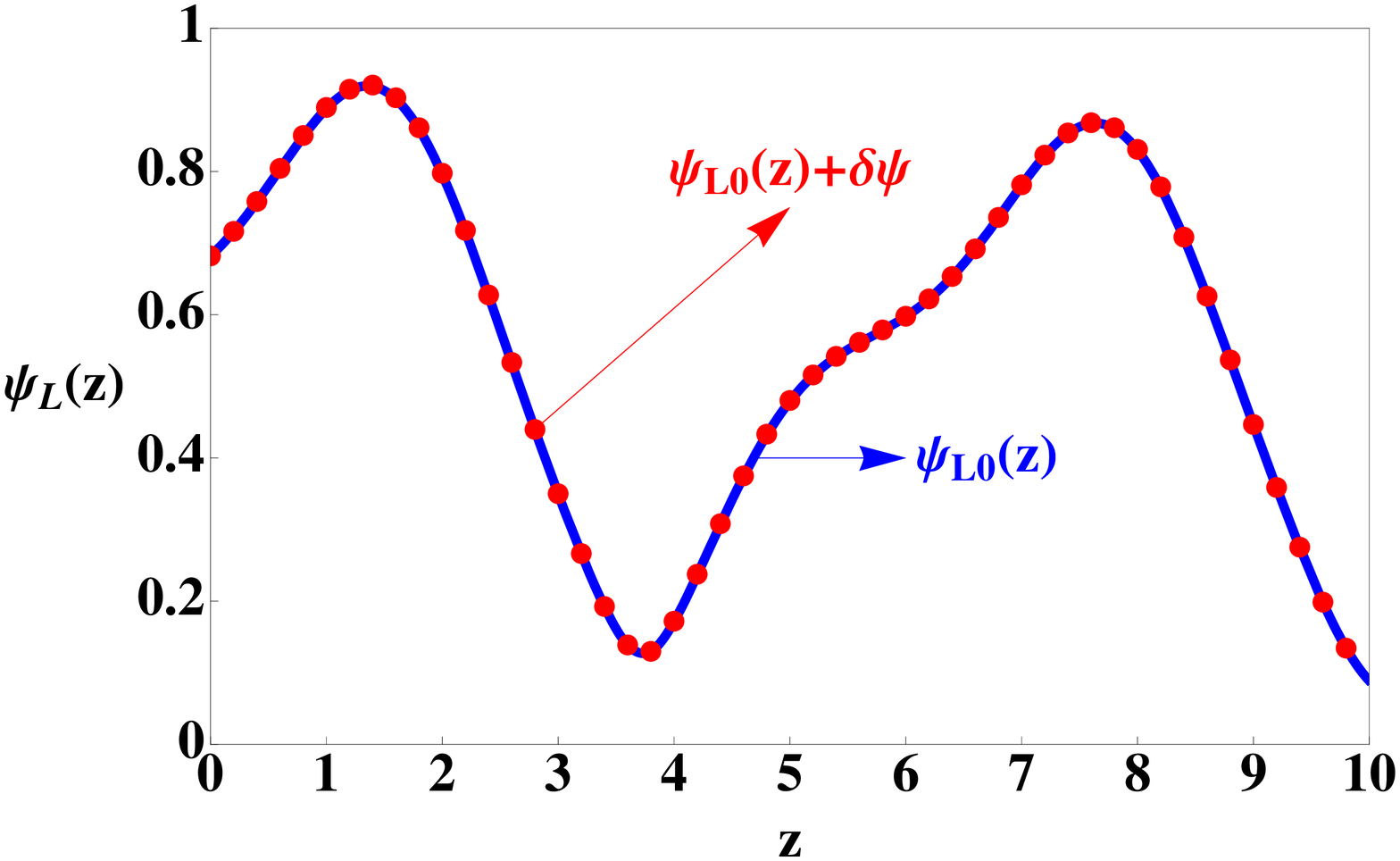}}
\caption{The solutions of the (a) upper, (b) middle, and (c) lower branches along with the perturbation is shown as a function of space. The blue (solid) lines represent the solutions of Eqs.~(\ref{CDE1}) and (\ref{CDE2}), whereas the red (dotted) lines show the behavior of the solutions along with the perturbation. Only the lower branch is stable.
}
\label{soln}
\end{figure*}	

\section{Linear stability analysis}
So far, we have focused on the presence of multiple-valued transmission peaks, since it is a common phenomenon in externally driven nonlinear systems.  At certain parameter regime we have found that a `duck-head' feature appears, which arises due to the interplay between the nonlinearity and the periodic lattice potential. 
More generally, three transmission points may appear in the transmission spectra for sufficiently strong nonlinearity, which we denote as the upper ($U$), middle ($M$) and lower ($L$) branch. A previous time-dependent-driving study in a similar BEC setting suggests that  only the lower branch is stable \cite{BNT1,paul}. We performed a linear stability analysis to check the stability of all three branches in the present system by the following method. Imagine applying a small perturbation ($\delta \psi \ll 1$) to the obtained stationary solutions $\psi_{I0}$ : $\Psi_{I}(z,t) = (\psi_{I0} (z)+ \delta \psi(z,t)) e^{- i \epsilon t}$, where $I = U$, $M$ and $L$.  Now, substituting this in Eq.~(\ref{LL}) and neglecting higher order terms, one obtains the equation in terms of the perturbation $\delta \psi$. 
\begin{eqnarray}
i \frac{\partial \delta \psi}{\partial t} &= &- \frac{1}{2 m} \frac{\partial^{2} \delta \psi}{\partial z^{2}} + (2 V \cos^{2} (z) - \epsilon) \delta \psi \nonumber \\  &&+ 4 \chi |\psi_{I0}|^{2} \delta \psi +2\chi \psi^{2}_{I0} \delta \psi.
\label{ptb}
\end{eqnarray}

Figure \ref{soln} illustrates the stationary solutions of Eqs.~(\ref{CDE1}) and (\ref{CDE2})  and the perturbed solutions found from Eq.~(\ref{ptb}) for the three branches. In our numerical calculation, we have applied a small Gaussian profile as an initial perturbation and considered the following boundary conditions : $\delta \psi(0,t) - i \frac{l_{coh} m_{R}}{m} \partial_{z}\delta \psi(0,t) = 0$ and $\delta \psi(d,t) + i \frac{l_{coh} m_{R}}{m} \partial_{z}\delta \psi(d,t) = 0$.  It can be clearly seen from Figs.~\ref{soln}(a) and (b)  that the solutions of the upper and middle branches are not linearly stable. 
We have further calculated the quantity, 
\begin{equation}
\Xi = \frac{\ln\{Re[\delta \psi(z,t+\delta t)]\} - \ln\{Re[\delta \psi(z,t)]\}}{\delta t},
\end{equation}
which in the limit of $\delta t \rightarrow 0$ and $t\rightarrow \infty$, approaches the largest eigenvalue of the perturbation mode evolving under Eq.~(\ref{ptb})  \cite{wright,akhmediev}.  
Explicit calculations show that the eigenvalues for the upper and middle branches are positive in the steady state and therefore unstable. 

We therefore conclude that the resonant transport will be suppressed during the propagation in the presence of a strong nonlinearity. This is in good agreement with the findings reported in \cite{BNT1,paul}.  It may, however, be possible to stabilize the upper branch to experimentally access it by applying a temporal modulation in the potential, as shown in \cite{BNT1} for a double well potential. We leave this for a future work.

\section{Summary and conclusion}
We have introduced a slow light system exhibiting EIT nonlinearities that provides an interesting platform for investigating linear and nonlinear out-of-equilibrium physics under the influence of a potential. We have shown that through varying certain optical parameters, such as single photon detunings, one can probe linear and nonlinear regimes with or without the lattice potential and thereby observe interesting phenomena in the transmission spectrum. Moreover, we have observed that a `duck-head'-like feature appears at the edge of the bandgap that results from the interplay between the lattice potential and nonlinearity. We have also performed a linear stability analysis in order to show which branches of the transmission spectra can be populated.

\acknowledgments
We would like to acknowledge the financial support provided by the National Research Foundation and Ministry of Education, Singapore. We thank Dr.~B.~M.~Rodr\'iguez-Lara for many helpful discussions, especially at the early stage of the work. P.~D.~thanks Dr.~J.~Bandyopadhyay for many useful discussions.

\end{document}